\documentclass[12pt]{article}
\usepackage{latexsym}
\usepackage{amssymb}
\usepackage{epsfig,amsmath,graphics}
\usepackage{multicol,bbm}

\newfont{\kreuz}{msbm10 scaled\magstep1}
\newfont{\Deutsch}{eufb10 scaled\magstep1}
\newfont{\deutsch}{eufb10}
\newfont{\schreib}{eusm10 scaled\magstep1}
\begin{document}

\newcommand{\be}{\begin{equation}}
\newcommand{\ee}{\end{equation}}
\newcommand{\ba}{\begin{array}}
\newcommand{\ea}{\end{array}}
\newcommand{\bea}{\begin{eqnarray}}
\newcommand{\eea}{\end{eqnarray}}
\newcommand{\bma}{\begin{matrix}}
\newcommand{\ema}{\end{matrix}}
\newcommand{\bpm}{\begin{pmatrix}}
\newcommand{\epm}{\end{pmatrix}}
\newcommand{\nn}{\nonumber}

\begin{titlepage}
\vspace*{0.2in}
\begin{center}
{\large\bf Renormalization Group Fixed Point with a Fourth Generation: Higgs-induced Bound States and Condensates}
\end{center}
\vspace{0.2in}

\begin{center}
P.Q. Hung and Chi Xiong \\
~~\\
pqh@virginia.edu,~xiong@virginia.edu \\
 Department of Physics, University of Virginia,\\
 Charlottesville, Virginia 22901\\
\end{center}

\begin{abstract}

In the Standard Model with four generations, the two-loop renormalization group equations for the Higgs quartic 
and Yukawa couplings show a quasi fixed point structure which does not appear at the one-loop level. This quasi
fixed point behavior indicates a possible restoration of scale symmetry above some physical cut-off scale $\Lambda_{FP}$. We conjecture that there exists a true fixed point which is reached at a similar energy scale. 
If the masses of the fourth family are sufficiently large, this cut-off scale, 
$\Lambda_{FP}$, is situated in the range of a few TeV to the order of $10^2$ TeV, above which the Higgs quartic and 
Yukawa couplings become practically constant. We found that around $\Lambda_{FP}$ the strong Yukawa couplings make 
it possible for the fourth generation to form bound states, including composite extra Higgs doublets. In this scenario the 
fourth generation condensates are obtained without introducing Technicolor or other unknown interactions. 
\end{abstract}

\end{titlepage}

\newpage

\section{Introduction}

~~~~~~~

One of the minimal extensions of the Standard Model (SM) is the fourth generation of quarks and leptons 
\cite{Kribs:2007}-\cite{Buras:2010pi}. It is known that precision data do not exclude the existence of the fourth generation 
\cite{Kribs:2007, Polonsky, Novikov:2002, Chanowitz:2001}. Besides many interesting physical applications it has brought 
(for a review, see \cite{Frampton}), the fourth generation also leads to new approaches in solving theoretical problems 
of the Standard Model with three generations, for example,  
the mechanism for dynamical symmetry breaking of the electroweak symmetries by condensates of the fourth generation quarks and
leptons \cite{Hill1990}, the improvement of the convergence of the three SM gauge couplings due to the Yukawa coupling 
contributions from the fourth generation \cite{Hung:1997zj}, the possibility of electroweak baryogenesis through first-order
phase transition of the SM with four generation \cite{Son}, the dimensional analysis of the CP violation based on the
Jarlskog invariants generalized to four generations \cite{Hou} etc.  

In this paper we study the quasi fixed point structure of the two-loop renormalization group equations (RGEs) for the Higgs quartic 
and Yukawa couplings in the SM with four generations, and its physical consequences. 
(In \cite{DESB_HX} we continue the discussion of condensate formation by performing a non-perturbative analysis using the Schwinger-Dyson equation in the ladder approximation.)
We will consider relatively heavy fourth family, more precisely 250 to 500 
GeV for the mass of the fourth quark, which is becoming an interesting parameter region since the experiments 
searching for the fourth family, for instance, the CDF AALTONEN ($p\bar{p}$ at 1.96 GeV), had put the mass of the fourth 
generation to be above 338-385 GeV \cite{PDG}. (We consider in this paper also values of the 4th quark mass lower than the experimental bounds simply as a matter of comparison and illustration.) Such ultraviolet quasi fixed points were previously
found and studied in the context of SU(5) gauge coupling unification \cite{Hung:1997rw} with values comparable to those found
in the present paper. The emphasis in this paper is however on the formation of bound states near the fixed
point and on the possibility of the restoration of scale symmetry along with its physical implications. Based on the existence of the quasi fixed point at the two loop level, we conjecture that there is a true fixed point which appears at the same energy scale as the two-loop quasi fixed point, albeit possibly at a lower value of the Yukawa coupling.

With such large initial masses of the fourth family, the evolutions of the Higgs quartic and Yukawa couplings at two-loop level are significantly different from the cases with smaller masses. A non-trivial (quasi) fixed point is found by solving the equations $ \beta_Y =0 $, 
where $\beta_Y$ are the two-loop $\beta$-functions for Higgs quartic coupling and Yukawa couplings of the top quark and the fourth 
generation.
The energy scale where they increase rapidly and reach the fixed point, denoted by $\Lambda_{FP}$, depends on the 
fourth quark mass $m_q$ and the fourth lepton mass $m_l$, especially $m_q$. For example, for $m_q =120 $ GeV and 
$m_l =100 $ GeV, $\Lambda_{FP}$ can be as large as $\sim 10^{16}$ GeV, while $m_q =500 $ GeV and $m_l =400 $ GeV can 
bring $\Lambda_{FP}$ down to less than 10 TeV. Since the new scale $\Lambda_{FP}$ signals a quasi fixed point, instead of a Landau 
pole that one can find at the one-loop level, it is natural to further study and explore new phenomena around and beyond $\Lambda_{FP}$.
Above $\Lambda_{FP}$ the masses of Higgs, top quark and fourth family are of order $\sim 1$ TeV and the result is quite 
impressive.
Firstly, the quasi fixed point could provide a natural physical cutoff  $\Lambda_{FP}$ to the Standard Model because the possible restoration of scale symmetry 
could have several implications. 
For instance, above $\Lambda_{FP}$, the couplings of Higgs-Yukawa sector stay practically constant and this property may give a hint on a possible link to
the hierarchy problem \cite{HX}.
Secondly, at $ \Lambda_{FP} $ the Yukawa couplings become strong enough for the fourth family to form bound states,
without introducing Technicolor 
\cite{TC} or other unknown interactions. When the minimum of the quartic coupling vanishes, the Yukawa interaction
by exchanging the Higgs scalar becomes infinitely long range. Such a scenario gives rise to the formation of Higgs-like 
condensates of the fourth family which will contribute to the symmetry breaking of the SM. 
Thirdly, such bound states, together with the fundamental Higgs doublet, can be effectively applied to the phenomenology as 
multiple Higgs doublets. This hybrid (fundamental plus dynamical) Higgs mass spectrum will give rise to a rich phenomenology 
which will be relevant to the Large Hadron Collider (LHC) and even to the International Linear Collider (ILC).

This paper is organized as follows: In Sec. II we present both analytic and numerical analyses of the two-loop renormalization 
group equations (RGEs) for the Higgs quartic and Yukawa couplings and the gauge couplings. We solve the vanishing 
$\beta$-function equations and show that the root corresponds to a quasi fixed point of the Higgs quartic and Yukawa couplings, 
and it can be reached at a TeV scale if the initial masses of the fourth family are large enough. The differences between 
the light and the heavy mass cases are discussed. Sec. III deals with bound state formation around $\Lambda_{FP}$. It includes 
the condition for bound state formation and a discussion of Higgs-like condensates.
Conclusions and discussions are given in Sec. IV.

\section{Two-loop RGEs and Quasi Fixed Point Structure}
~~~~

In subsection 2.1 we study the two-loop RGEs for the quartic and Yukawa couplings and gauge couplings and show that 
there exists a non-trivial quasi fixed point in the Yukawa sector by solving the vanishing two-loop $\beta_Y$ function equations, 
and consequently, a natural scale $\Lambda_{FP}$ is introduced. 
The relation between the RG evolutions of the Yukawa sector and gauge sector is also discussed. We then perform a direct 
integration of these RGEs by a numerical approach and the results are given in subsection 2.2.

\subsection{Two-loop $\beta$ function and Quasi Fixed point}
~~~~
In the Standard Model with four generations, at two-loop level,
the renormalization group equations are given by \cite{Hung:1997zj, twoloop2} :
\be \label{RGE}
16 \pi^{2} \frac{dY}{dt} = \beta_Y
\ee
where $Y$ represents the quartic coupling $\lambda$, the Yukawa couplings
$g_t^2$, $g_q^2$, $g_l^2$ and gauge coupling constants $g_i^2, i=1,2,3$ respectively, and
\begin{eqnarray} \label{lambda}
\beta_{\lambda} =&& 24 \lambda^{2} + 4 \lambda( 3 g_{t}^{2}+6 g_{q}^{2} + 2 g_{l}^{2}-2.25 g_{2}^{2}-0.45 g_{1}^{2})\nonumber \\
&&-2( 3 g_{t}^{4} + 6 g_{q}^{4} + 2 g_{l}^{4})+(16 \pi^{2})^{-1}[30 g_t^{6}\nonumber \\
&&+48 g_q^{6}+ 16 g_l^{6} -(3 g_t^{4} - 80 g_3^{2} (g_t^{2}+ 2 g_q^{2}))\lambda \nonumber \\
&&-6\lambda^{2} (24 g_t^{2} + 48 g_q^{2} + 16 g_l^{2})-312\lambda^{3}\nonumber \\
&&-32 g_3^{2}( g_t^{4} + 2 g_q^{4})]
\end{eqnarray}
\begin{eqnarray}
\beta_{g_t} &=& g_t^{2} \{9 g_t^{2} +12 g_q^{2} + 4 g_l^{2}-16 g_3^{2}-4.5 g_2^{2}-1.7 g_1^{2}\nonumber \\
&&+(8 \pi^{2})^{-1}  [1.5 g_t^{4}-2.25 g_t^{2}(6 g_q^{2}+ 3 g_t^{2}+ 2 g_l^{2})\nonumber \\
&&-12 g_q^{4}- (27/4) g_t^{4} - 4 g_l^{4}+ 6 \lambda^{2} \nonumber \\
&&+g_t^{2}(-12 \lambda + 36 g_3^{2}) +40 g_q^2 g_3^2-(892/9) g_3^{4}] \}
\end{eqnarray}
\begin{eqnarray}
\beta_{g_q} &=& g_q^{2} \{6 g_t^{2} +12 g_q^{2} + 4 g_l^{2}
-16 g_3^{2}-4.5 g_2^{2}-1.7 g_1^{2}\nonumber \\
&&+(8 \pi^{2})^{-1}  [3 g_q^{4}-g_q^{2}(6 g_q^{2}+ 3 g_t^{2}+ 2 g_l^{2})\nonumber \\
&&-12 g_q^{4}- (27/4) g_t^{4} - 4 g_l^{4}+ 6 \lambda^{2} \nonumber \\
&&+g_q^{2}(-16 \lambda + 40 g_3^{2}) + 20 g_t^2 g_3^2 -(892/9) g_3^{4}] \}
\eea
\bea
\beta_{g_l} &=& g_l^{2} \{6 g_t^{2} +12 g_q^{2} + 4 g_l^{2}
-4.5 (g_2^{2}+ g_1^{2})\nonumber \\
&&+(8 \pi^{2})^{-1}  [3 g_l^{4}-g_l^{2}(6 g_q^{2}+ 3 g_t^{2}
+ 2 g_l^{2}) -12 g_q^{4}\nonumber \\
&&- (27/4) g_t^{4} - 4 g_l^{4}+ 20 g_t^2 g_3^2 + 40 g_q^2 g_3^2 \nonumber\\
&& + 6 \lambda^{2} -16\lambda g_l^{2}] \}
\eea
\bea  \label{g1}
\beta_{g_1} &=& g_1^{4} \{ (163/15)+(16 \pi^{2})^{-1}[
(787/75) g_1^{2} + 6.6 g_2^{2}\nonumber \\
&&+(352/15) g_3^{2}-3.4 g_t^{2}-4.4 g_q^{2}-3.6 g_l^{2}] \}
\eea
\bea  \label{g2}
\beta_{g_2} &=& g_2^{4} \{ -(11/3)+(16 \pi^{2})^{-1}[
(11/5) g_1^{2} + (133/3) g_2^{2}\nonumber \\
&&+32 g_3^{2}-3 g_t^{2}-6 g_q^{2}-2 g_l^{2}] \}
\eea
\bea  \label{g3}
\beta_{g_3} &=& g_3^{4} \{ -(34/3)+(16 \pi^{2})^{-1}[
(44/15) g_1^{2} + 12 g_2^{2}\nonumber \\
&&-(4/3) g_3^{2}-4 g_t^{2}-8 g_q^{2}]\}.
\eea

In the above equations, we have taken the same assumptions as in \cite{Hung:1997zj}:
the fourth family has a Dirac neutrino mass and both quarks and leptons are degenerate $SU(2)_L$ doublets, with Yukawa 
couplings denoted by $g_q$ and $g_l$ respectively. For simplicity, in solving the RGEs, the mass difference between the fourth 
up-type quark $U$ and the down-type quark $D$ is neglected, although it might be interesting in the data-fitting of $S$ and $T$ 
parameters. Also, in the evolution of $\lambda$ and the Yukawa couplings, we have neglected,
in the two loop terms, contributions involving $\tau$ and bottom Yukawa couplings as well as those of the
lighter fermions.

We now study the quasi fixed-point structure (at the two-loop level) of the RGEs (\ref{RGE}) (see remarks made in
\cite{Hung:1997zj} concerning ultraviolet fixed points and \cite{Hung:1997rw} where such fixed points were used
in the context of SU(5) gauge coupling unification).
This suggests vanishing $\beta$ functions for the Yukawa couplings and furthermore, a vanishing $\beta$
function for the quartic coupling
\be \label{fixedpoint}
\beta_{Y} |_{{g_{1,2,3}=const.}} = 0, ~~\textrm{for}~~Y= \lambda, g_t^2, g_q^2, g_l^2.
\ee
\begin{table}[!tbp]
\centering
\begin{tabular}{c|c|c|c|c|c|c|c}
\hline
\hline
$g_3^2$  & $g_2^2$ & $g_1^2$ && $\lambda$  & $g_t^2$ & $g_q^2$ & $g_l^2$ \\
\hline
1.478 & 0.425 & 0.213 &&  17.561 & 31.407 & 52.298 & 56.583 \\
1.225 & 0.413 & 0.217 &&  17.457 & 31.200 & 52.185 & 55.664 \\
1.003 & 0.404 & 0.223 &&  17.376 & 31.073 & 52.147 & 54.934 \\
0.902 & 0.396 & 0.226 &&  17.339 & 31.014 & 52.126 & 54.604 \\
0.815 & 0.386 & 0.230 &&  17.308 & 30.963 & 52.107 & 54.321 \\
0.751 & 0.381 & 0.232 &&  17.285 & 30.925 & 52.091 & 54.113 \\
0.652 & 0.366 & 0.239 &&  17.249 & 30.866 & 52.066 & 53.792 \\
0.565 & 0.354 & 0.245 &&  17.218 & 30.814 & 52.042 & 53.511 \\
0.457 & 0.330 & 0.260 &&  17.180 & 30.748 & 52.011 & 53.162 \\
0.304 & 0.284 & 0.304 &&  17.125 & 30.655 & 51.966 & 52.661 \\
0.999 & 0.666 & 0.333 &&  17.339 & 31.039 & 52.089 & 54.817 \\  
0.500 & 0.500 & 0.500 &&  17.164 & 30.754 & 51.990 & 53.152 \\
0.000 & 0.000 & 0.000 &&  17.059 & 30.488 & 51.902 & 51.902 \\
\hline
\hline
\end{tabular}
\caption{{\small The first three columns list different values of gauge couplings  
(except for the imaginary case in the last row where the gauge couplings are turned off). The last four columns list the 
roots of the vanishing $\beta$-function equations for the Higgs quartic and Yukawa couplings.}}
\label{fptable}
\end{table}
The solutions to (\ref{fixedpoint}) with different gauge couplings are listed in Table \ref{fptable}.
In Table \ref{fptable} some of values of gauge couplings are chosen from the their RG evolutions, some are
just arbitrary numbers at the same order of magnitude,
plus the particular case where the gauge couplings are turned off (the last row of Table \ref{fptable}).
Equations (\ref{fixedpoint}) then become polynomial equations for $\lambda, g_t^2, g_q^2$ and $g_l^2$.
They can be solved numerically and the only solution with all roots being positive is listed in the above table. For example,
at $E \sim 50$ TeV, the gauge couplings are approximately $g_1^2 = 0.230,~g_2^2 =0.386, ~g_3^2 =0.815$ and solving
(\ref{fixedpoint}) yields $\lambda= 17.308, ~g_{t}^2 = 30.963,~g_{q}^2 =52.107, ~g_{l}^2 =54.321 $.
The average values of the quasi fixed points of the quartic and Yukawa couplings are approximately
\be \label{fpvalues}
{\lambda}^{*} \approx 17, ~g_{t}^{2*} \approx 31,~g_{q}^{2*} \approx 52, ~g_{l}^{2*}\approx 54,
\ee
or, using a more conventional notation with $\alpha_i \equiv g_{i}^2/4 \pi$,
\be \label{fpvalues2}
\alpha_{\lambda}^{*} \approx 1.35, ~\alpha_{t}^{*} \approx 2.47,~\alpha_{q}^{*} \approx 4.14, ~\alpha_{l}^{*}\approx 4.3,
\ee
which correspond to $\overline{MS}$ masses (using $\overline{m}_H =  v\sqrt{2 \lambda} $ and $\overline{m}_f = v g_f /
\sqrt{2},~v=246$ GeV )
\be \label{fpmass}
\overline{m}_H^{*} = 1.44~\textrm{TeV}, \overline{m}_t^{*} = 0.97 ~\textrm{TeV}, 
\overline{m}_q^{*} = 1.26 ~\textrm{TeV}, \overline{m}_l^{*} = 1.28 ~\textrm{TeV}.
\ee
Note that our analysis depends solely on the values of the couplings at some initial energy scale. However, for the clarity
of the {\em illustration}, we translate these couplings into {\em naive} masses in units of the electroweak scale $ v=246$ GeV.
We have to keep in mind that the inclusion of the dynamical extra Higgs doublets, which will be discussed in Sec. III, will
modify the {\em naive} masses to the {\em physical} masses \cite{HX, HX2}.
Also, the quasi fixed point values shown in (\ref{fpvalues}) are comparable to the ones found in \cite{Hung:1997rw} where the emphasis
was however on SU(5) gauge coupling unification.

For comparison and in order to show the significance of the scale $\Lambda_{FP}$, we show in Fig.\ref{Landau} the evolution
of the couplings of the Higgs-Yukawa sector at the one and two-loop levels for comparison. At one loop, we notice the
appearance of the Landau singularity at an energy scale similar to that of the quasi fixed point. Only for ``light''
fourth generations fermions (a case which is ruled out experimentally anyway) do the two scales differ. As can be seen
from Fig.\ref{Landau}, the couplings in the Higgs-Yukawa sector increase in values before approaching the quasi fixed point values.
As we will discuss in Sec. III, the dynamics of bound states and condensates occur at values of the couplings which are
smaller than those of the quasi fixed point.

At this point, a few words concerning the magnitudes of the couplings in (\ref{fixedpoint}) and (\ref{fpvalues}) are in
order. The values shown appear to be ``large'' and this may trigger a natural question as to the validity of perturbation
theory when the evolution of the couplings approaches these quasi fixed points. The relevant question is the form
of the expansion parameters in the $\beta$-functions. For instance, for the case of a single coupling and with
$\alpha= g^2/4\pi$, the $\beta$-function can be written as $\beta(g)=g(\beta_0(\alpha/4\pi)+ \beta_1(\alpha/4\pi)^2+...$,
with $dg/dt = \beta(g)$ (see e.g. \cite{abbott}). The expansion parameter is $\alpha/4\pi= g^2/16\pi^2$. 
Consequently, it is reasonable to expect the expansion parameters in our case
to be $\lambda/16\pi^2$, $g_t^2/16\pi^2$, $g_q^2/16\pi^2$ and $g_l^2/16\pi^2$. Even at the quasi fixed point (\ref{fpvalues}),
one expects $\lambda^{*}/16\pi^2 \approx 0.11$, $g_t^{2*}/16\pi^2 \approx 0.2$, $g_q^{2*}/16\pi^2 \approx 0.33$, 
$g_l^{2*}/16\pi^2 \approx 0.34$. The values of the expansion parameters as the couplings approach the
fixed point are smaller than the aforementioned values and it is not unreasonable to suspect that
the two-loop quasi fixed point might be of the same order as the true fixed point. 
As we have mentioned in the Introduction, a non-perturbatiive analysis of condensate formation using the Schwinger-Dyson equation in the ladder approximation as done in \cite{DESB_HX} leads to the existence of a critical coupling $\alpha_c = \pi/2 \approx 1.57$ above which condensate formation can be realized. This is about a factor of three smaller than the quasi fixed point values listed (\ref{fpvalues2}).  What \cite{DESB_HX} shows is that condensate formation can occur at a value of the couplings (i.e. $> \alpha_c = \pi/2 \approx 1.57$) which can be much lower (e.g. $\sim 2$) than those of the quasi fixed points, and hence more manageable. In fact, as one can see from Fig. (\ref{Landau}), for a heavy fourth generation the physical cutoff scales coming from the RGE at one loop and at two loops are basically the same and of $O(TeV)$. Furthermore, the Yukawa coupling with a value slightly above the critical coupling is the same whether one uses the RGE at one loop or two loops. In consequence, it might be reasonable to expect that this will not change when three or more loops (if one knew how to calculate it) are considered. It is also tempting to speculate that the true fixed point might be located at a similar value of the coupling where condensate formation occurs. This is our conjecture.
An important point that we made in the above paragraph and would like to repeat here is:
the dynamics of bound states and condensates occur at values of the couplings which are
smaller than those of the quasi fixed point. Notice that the values of the couplings at the quasi fixed point are
not representative of the dynamics of condensates and bound states. This point will be further clarified
in Sec. III.


The next issue we would like to investigate is the stability of the quasi fixed points shown in (\ref{fixedpoint})
and (\ref{fpvalues}). If they are proven to be ``substantially'' stable, it is plausible (our conjecture) that the true
ultraviolet (UV) fixed point might not be too different from the two-loop quasi fixed points found in this paper.
In particular, it is also important and interesting to see how ``attractive'' these stable (quasi) fixed
points turn out to be.
(As shown numerically in the next section, different initial values for the quartic
and 4th-generation Yukawa couplings at the electroweak scale all lead to the same UV quasi
fixed points similar to the above values, albeit at different energy scales.)  Mathematically speaking, the RG flow
into the fixed point is governed by the eigenvalues of the stability matrix which depends on
the gradients of the $\beta$-functions. 
Here, the stability matrix is a 4x4 matrix whose elements are given by
\be
M^i_{~j} \equiv \left( \frac{\partial \beta^i(Y)}{\partial Y^j} \right)_{Y=Y^*}, ~~~~~ i,j=1,2,3,4
\ee
and $Y^{1,2,3,4}= (\lambda, g^2_t, g^2_q, g^2_l)$ respectively. As we have shown above in (\ref{fixedpoint}),
the fixed points of the Higgs-Yukawa sector are relatively insensitive to the gauge coupling sector.
For this reason, we concentrate on the Higgs-Yukawa sector alone in our investigation of the
fixed point stability question.
Whether the UV quasi fixed point (\ref{fpvalues})
is stable or not depends on the signs of the eigenvalues of $M^i_{~j}$. Evaluating $M^i_{~j}$ at (\ref{fpvalues}), 
we found its eigenvalues to be $(-3951.67, -1143.45, -428.86, -329.73)$, which are all negative. The
corresponding eigenvectors are
\bea  
\nonumber
 &&\{[0.98,0.03,0.13,0.14],[-0.31,-0.38,-0.59,-0.64],\\
 &&[-0.03,0.98,-0.12,-0.13],[0.001,0.04,-0.35,0.93]\}.
\eea
The negative signs of {\em all} the eigenvalues indicate that the (quasi) fixed point (\ref{fpvalues}) is {\em UV stable}
and {\em attractive} in all four directions. (An eigenvalue with a positive sign would
indicate that the fixed point is repulsive.) 
Furthermore, the {\em large magnitude} of the eigenvalues indicate that this UV fixed point is highly {\em attractive}.
Heuristically speaking, if the $\beta$-functions were like the ``potential'' and the couplings were like ``coordinates'',
the stability matrix eigenvalues (the gradients of the $\beta$-function) would be analogous to the ``forces''. The
``large'' eigenvalues would be analogous to ``strong attractive forces'' pulling the couplings toward
their (quasi) fixed points. It is also possible that the large eigenvalues ``protect'' the RG flow against small fluctuations
in the values of the quasi fixed points.
Using other sets of quartic and Yukawa couplings from Table 1, we obtain similar results.  
For example, using the last row of Table 1, we obtain the eigenvalues $(-3956.13, -1082.08, -403.31, -307.78) $.
Furthermore let us switch on the gauge interactions and consider the first row of Table 1. The corresponding
eigenvalues of the stability matrix are $(-4263.18, -1208.24, -461.05, -370.70)$. One may try other cases
and the eigenvalues basically interpolate between these two extreme cases. 
In summary, all eigenvalues of the stability matrix are {\em negative} in our case, which shows 
that around the fixed point (\ref{fpvalues}), there is a finite region where {\em all} RG flows are attracted 
to the fixed point. Therefore (\ref{fpvalues}) is a truly stable (quasi) fixed point. This will be confirmed
in the next section by numerically integrating the RGEs with different initial values.

An estimate of the values of the anomalous dimensions of various operators in the Higgs-Yukawa sector reveals
that the scalar quartic interaction term becomes more irrelevant as compared with  the Yukawa interaction term as
the energy increases. From Fig. (\ref{Yukawa}) shown in the next section,  the values of the couplings at $t=2.75$ corresponding
to $E \approx 1.4 \, TeV$ are $\lambda \sim 5$, $g_t^2 \sim 2.1$, $g_q^2 \sim 19.3$ and $g_l^2 \sim 14.6$. The 
values of the anomalous dimensions for the Yukawa and quartic interactions are estimated to be $\gamma_Y \sim 0.97$ and
$\gamma_\lambda \sim 14.77$. This little exercise indicates that  the Yukawa interactions are the dominant operators as
one approaches (from below) the TeV scale.  As we shall discuss below, it is these interactions which are responsible for
the formation of condensates which spontaneously break the electroweak symmetry.

We now estimate the contributions of the Yukawa couplings of the top quark and the fourth generation to the gauge sector.
These contributions enter in the $\beta$ functions of the gauge sector at the two-loop level, at the order of $g^2_f / 16 \pi^2$
compared to the one-loop $\beta$ functions. For the fixed points values given in (\ref{fpvalues}) suppressed by the loop factor
$16 \pi^2 $, these are about 
$g^{2*}_t / 16 \pi^2 \approx 0.2, ~g^{2*}_q / 16 \pi^2 \approx 0.33, ~g^{2*}_l / 16 \pi^2 \approx 0.34 $ plus 
$\lambda^{*}/ 16 \pi^2 \approx 0.11$. 
Plugging the fixed point values of the Yukawa couplings (\ref{fpvalues}) into 
the gauge sector of the RGEs (\ref{g1})-(\ref{g3}), one finds that the magnitudes of the two-loop 
contributions are about -3.2, -2.9 and -3.4 respectively, which are significant corrections but the RG running of the gauge couplings
is still dominated by the one-loop factors 163/15, -11/3 and -34/3.  
This shows that the RGEs for the Higgs quartic and Yukawa couplings are almost decoupled from the RGEs for the gauge
couplings above the scale $\Lambda_{FP}$: the large values of quartic and Yukawa couplings at the fixed point do not affect much 
the evolutions of the gauge couplings except for improving a bit the convergence of the running gauge couplings. 
Meanwhile the evolutions of the gauge couplings contribute very little to the Yukawa 
sector and do not drive the quartic and Yukawa couplings away from their fixed point values. 
Around $\Lambda_{FP}$, a whole new spectrum of bound states of fourth generation quarks and leptons get formed-
many of which carrying the SM quantum numbers- as we will show below
and it will not be justified to use the two-loop RGEs (\ref{RGE}) for the gauge sectors (as well as for the light fermion Yukawa sectors). 
(The gauge couplings may even be driven toward a fixed point in the vicinity of $\Lambda_{FP}$.)

If we denote generically the quasi fixed point values of the various couplings by $g^{*}$ obtained at the two-loop
level i.e. they are solutions to $\beta^{(0)}(g^{*})+ \beta^{(1)}(g^{*})=0$, where $\beta^{(0)}$ and $\beta^{(1)}$ are the
one and two-loop terms in the $\beta$ function respectively, the assumptions made here and in what follows are the
following: (a) The higher loop terms, $\beta^{(n \geq 2)}(g^{*})$, are ``small'' compared with either $\beta^{(0)}(g^{*})$
or $\beta^{(1)}(g^{*})$; (b) The possible inclusion of these unknown higher order terms will not shift
the fixed point values by a significant amount.

The values of the couplings at the quasi fixed point are ``large'' but at the same time ``not too large'' (see the
remarks made above concerning this issue). It is interesting to look at systems where couplings of that size might
occur and where one might learn something from them even though they might be quite different from the present model.
One of such systems is the one studied by
Wilson and Fisher \cite{Wilson}, using the $\epsilon$-expansion ($\epsilon=4-d$) to calculate the critical exponent
of systems such as the magnetization in a ferromagnet, which is described by a single scalar $\phi$ and the effective
Lagrangian contains interactions $-g_4 \phi^4 /4!, -g_6 \phi^6 /6!$ and etc, in addition to the mass term $ -g_2 \phi^2 /2$.
For $d=4-\epsilon$ the RGEs for $g_2$ and $ g_4$ (here we use Weinberg's formulations \cite{Weinberg})
\bea
\mu \frac{d}{d\mu} g_4(\mu) &=& - \epsilon~g_4(\mu) + \frac{3 g_4^2(\mu)}{16 \pi^2} + \mathcal{O}( g_4^3(\mu))  \\
\mu \frac{d}{d\mu} g_2(\mu) &=&  g_2(\mu)[-2 + \frac{g_4(\mu)}{16 \pi^2} + \mathcal{O}( g_4 (\mu)) ]
\eea
Solving the $\beta(g_4), \beta(g_2)=0$ equations, one finds a non-trivial fixed point at \cite{Weinberg}
\be \label{WFP}
g_4^{*} = \frac{16 \pi^2 \epsilon}{3}, ~~~ g_2^{*} =0.
\ee
The critical exponent $\nu$ is then given by the $\epsilon$-expansion $ \nu = 1/2 + \epsilon/12 + 7 \epsilon^2/162
-0.01904 \epsilon^3 + \mathcal{O}(\epsilon^4) $ \cite{Wilson, Weinberg}. For the physical value $\epsilon=1$ this three-loop
calculation yields $\nu=0.61$, while the one-loop calculation yields $\nu=0.58$, and the experimental value \cite{Weinberg}
is $\nu=0.63 \pm 0.04 $. Note that if we look at Eq. (\ref{WFP}),  $\epsilon=1$ actually corresponds to
$g_4^{*} = 16 \pi^2 / 3 \approx 52.64$ or $ g_4^{*} / 16 \pi^2 = 1/3 $, which is not a small expansion parameter.(Notice that
our quartic coupling expansion $\lambda^{*} /16 \pi^2  \approx 0.11 $ at the fixed point.)
Although this example may not be directly related to our case, it does show that 
some expansion in relatively large parameters can work well in critical phenomena.

Now the next question is the value of the scale $\Lambda_{FP} $ where the Yukawa sector reach the quasi fixed point and 
couples to the gauge sector in the way as described above. More precisely, given the root values as in equation (\ref{fpvalues}),
at which energy scale can the quasi fixed point be reached?  This of course depends on the initial values of all couplings and their 
corresponding evolutions of RGEs.

\subsection{Numerical Results}
~~~~

In this section, we first begin with a short discussion on experimental constraints on the 4th generation. As we have mentioned in the Introduction, it has been realized in recent years that electroweak precision data do not rule out the possible existence of a 4th generation \cite{Kribs:2007, Polonsky, Novikov:2002, Chanowitz:2001}. Of relevance to our analysis is the direct bounds on the 4th generation masses. For the quarks, direct bounds from the Tevatron give $m_{t'} > 338\, GeV$ and $m_{b'} > 385\, GeV$ \cite{PDG}. (As we have stressed above, these bounds were obtained under a certain assumption about the decays of $t'$ and $b'$. By relaxing these assumptions, one can possibly lower these bounds (see Hung and Sher in \cite{Kribs:2007}).) As we will show below, our favorite mass range is 400-500 GeV since the energy scale where the quasi fixed point appears is of O(TeV). We start below with masses that are below the Tevatron bounds {\em only} for illustrative purpose: These low masses would yield a physical cutoff scale that is much larger than O(TeV) and this runs into severe fine-tuning problems as emphasized in \cite{HX} and \cite{DESB_HX}.

We integrate the RGEs (\ref{RGE}) by the Runge-Kutta-Fehlberg method (RKF45) with general initial conditions, i.e. different mass
combinations for the fourth quarks and leptons. For the gauge coupling sector, at $ E=M_Z \sim 91.2 $ GeV, we take \cite{PDG}
\be
\hat{\alpha}_s(M_z) = 0.1176 \pm 0.0020, ~\textrm{sin}^2\theta_w (M_z) = 0.2312, ~\hat{\alpha}^{-1}(M_z)=127.909.
\ee
For initial masses we will restrict to the cases with $\overline{m}_H \leq 750 $ GeV and $\overline{m}_l \leq \overline{m}_q \leq 500$ GeV 
at the electroweak scale. This is because we want to start the RG running from the perturbative region of the quartic and Yukawa couplings.
Initial masses $\overline{m}_H > 750 $ GeV and $\overline{m}_q > 500$ GeV  (see \cite{Holdom:2006, Chanowitz2009} for example) are possible
but they are beyond the scope of this paper. Again these masses are translated from the quartic and Yukawa couplings according to
$\overline{m}_H =  v\sqrt{2 \lambda} $ and $\overline{m}_f = v g_f \sqrt{2},~v=246$ GeV and the remarks made after Eq. (\ref{fpmass}) 
should be kept in mind.

Cases for which the initial $m_q > 250$ GeV will be referred to as the {\it heavy case} and by the {\it light case} otherwise. To see the
difference between these two cases let us first look at the following two examples. Again notice that the light mass case is presented {\em only for
illustration}, without taking into account the experimental lower bound on the fourth quarks. As we have mentioned in the Introduction, the most recent bounds on the 4th quark masses from the Tevatron are as follows.

Initial values of these two examples are given as (in $\overline{\textrm{MS}}$ scheme)

(1) light mass cases:
\be  \label{light}
\overline{m}_H = 224~\textrm{GeV}, ~\overline{m}_t = 161~\textrm{GeV},~\overline{m}_q = 120~\textrm{GeV},~\overline{m}_l = 100~\textrm{GeV};
\ee

(2) heavy mass cases:
\be  \label{heavy}
\overline{m}_H = 506~\textrm{GeV}, ~\overline{m}_t = 161~\textrm{GeV},~\overline{m}_q = 350~\textrm{GeV},~\overline{m}_l = 250~\textrm{GeV}.
\ee
\begin{figure}[ht]
\centering
\begin{tabular}{cc}
    \includegraphics[scale=0.75]{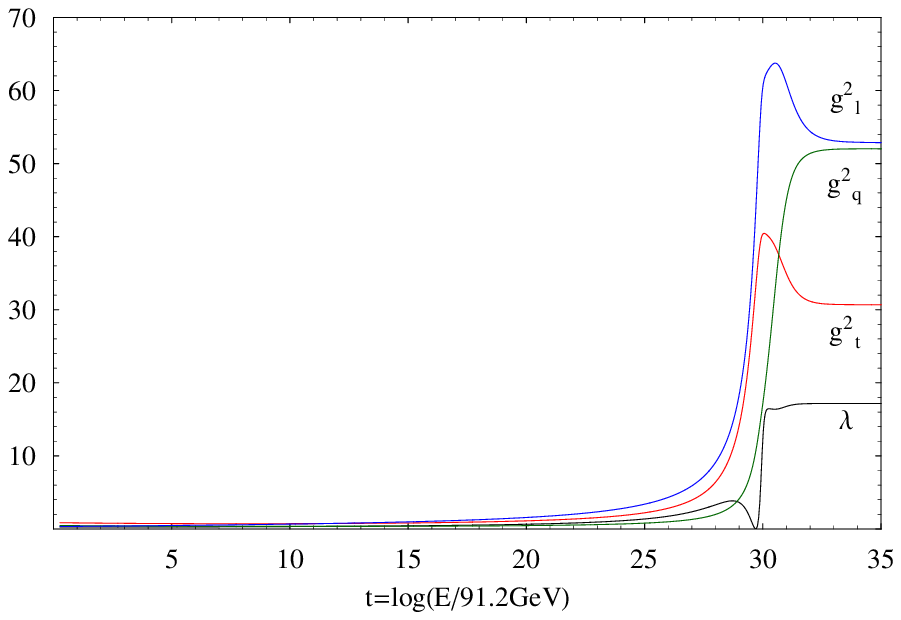} & \includegraphics[scale=0.75]{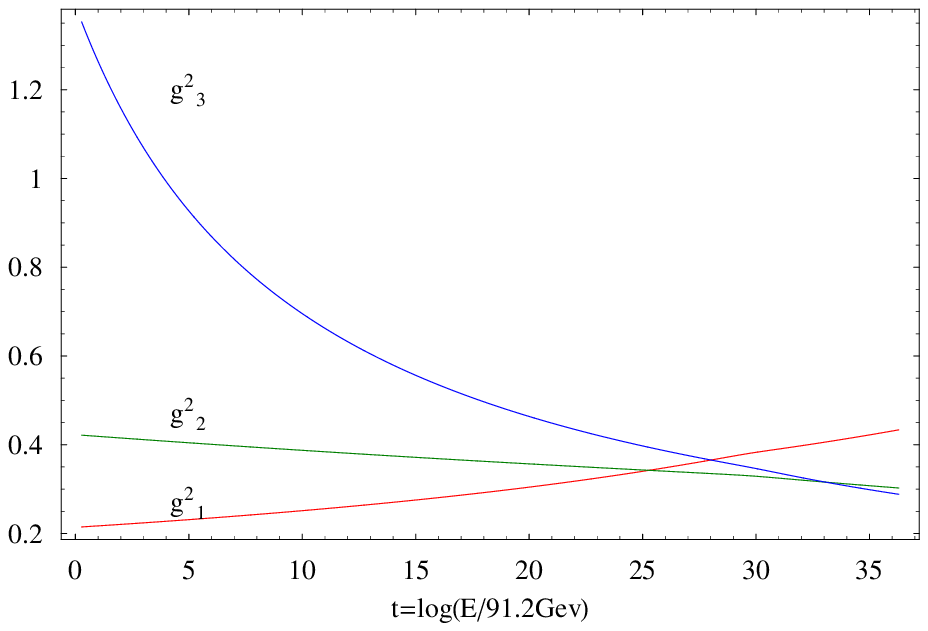}\\
    (a) & (b) \\
    \includegraphics[scale=0.75]{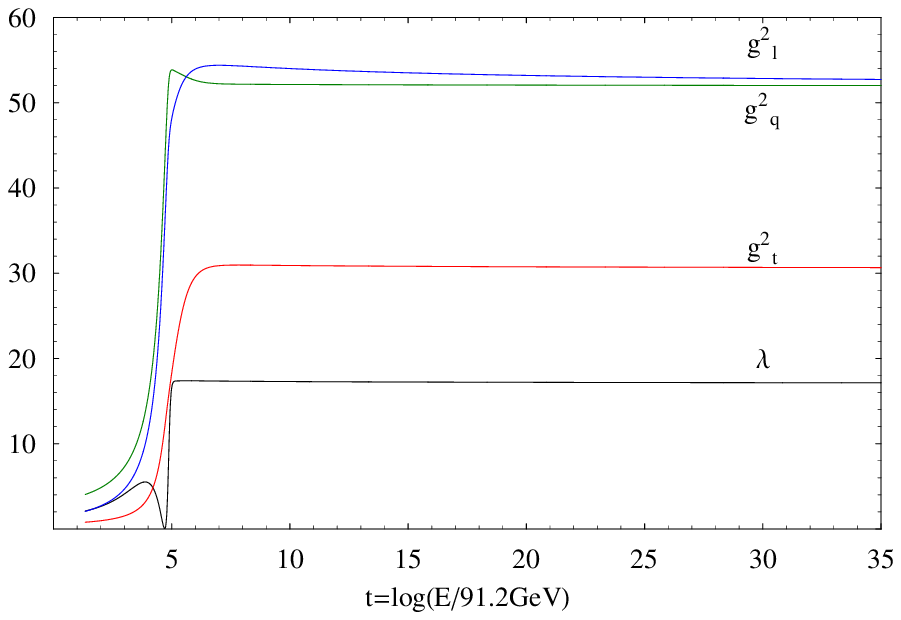} & \includegraphics[scale=0.75]{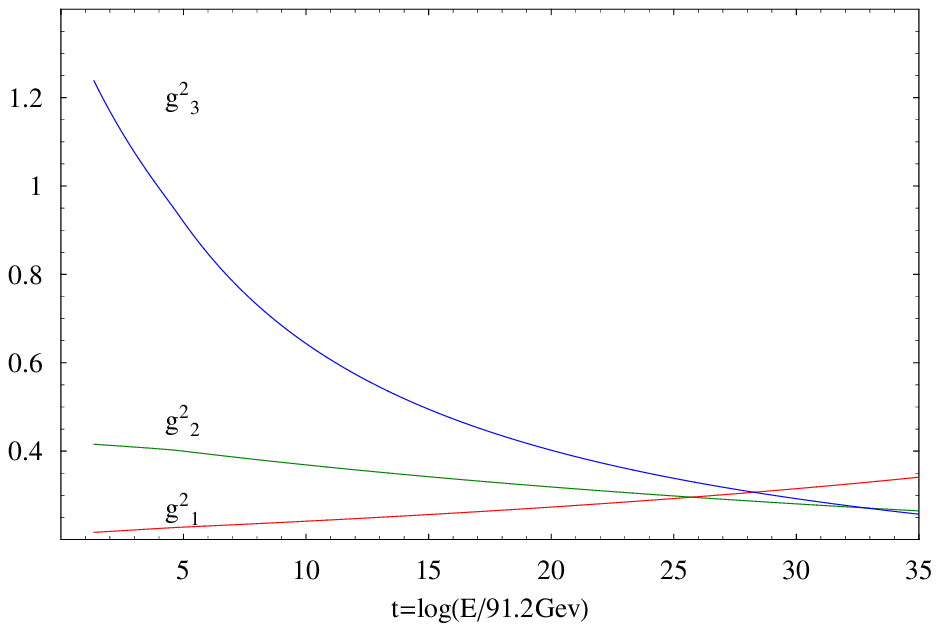} \\
    (c) & (d)
\end{tabular}
\caption{{\small The evolutions of the quartic and Yukawa coupling constants and gauge coupling
constants: figures (a), (b) are for the light mass case; figures (c), (d) are for the heavy mass case.}}
\label{light_heavy}
\end{figure}

From Fig.\ref{light_heavy} we can see that in the light mass case (1), the fixed point is not reached until $t \approx 32$ corresponding 
to the scale $\sim 10^{16} $ GeV, and the quartic and Yukawa couplings are running with small values in the perturbation region all the 
way to that scale. From Fig.\ref{light_heavy} (b) we also see that the gauge couplings evolve in a similar way to the case of Standard Model with three
generations. The heavy mass case (2) demonstrates a different scenario: the heavy masses of the fourth generation at the electroweak scale
drive the RGE flow to the fixed point around $t = 4 \sim 6 $, i.e., 5 TeV - 40 TeV. From the RG flows of the quartic and Yukawa couplings, 
the fixed point values are approximately
\be
\overline{m}_H^{\textrm{\tiny{FP}}} = 1.446 ~\textrm{TeV}, \overline{m}_t^{\textrm{\tiny{FP}}} = 0.965 ~\textrm{TeV}, \overline{m}_q^{\textrm{\tiny{FP}}} = 1.260 ~\textrm{TeV}, \overline{m}_l^{\textrm{\tiny{FP}}} = 1.282 ~\textrm{TeV}
\ee
which are in good agreement with the roots (\ref{fpvalues}) by solving the vanishing $\beta$ function equations $\beta_Y =0$. But the
more important and remarkable thing is that, by increasing their masses just about three times heavier,  the fourth generation fermions 
bring the fixed point scale $\Lambda_{FP}$ from $\sim 10^{16}$ GeV down to a few TeV.  Note that in Fig.\ref{light_heavy} (d) we assume that
the gauge couplings evolve according to the same RGEs above $\Lambda_{FP}$, which may change significantly if the contributions from new
dynamical degrees of freedom are included. As a result, Fig.\ref{light_heavy} (d) is given mainly for illustration since one does not expect
the two-loop RGEs for the gauge couplings to be valid above $\Lambda_{FP}$.
\begin{figure}[!tbp]
\centering
    \begin{tabular}{c}
    \includegraphics[scale=1.2]{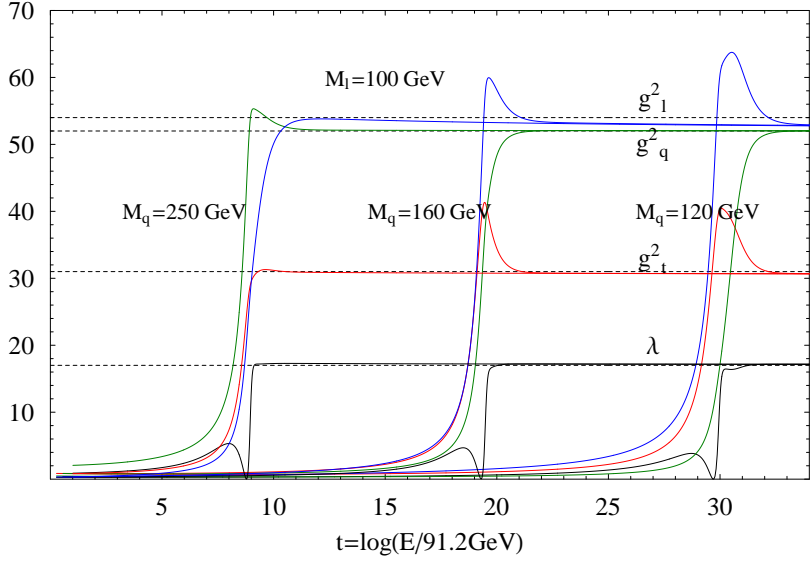}\\
    (a)\\
    \\
    \includegraphics[scale=1.2]{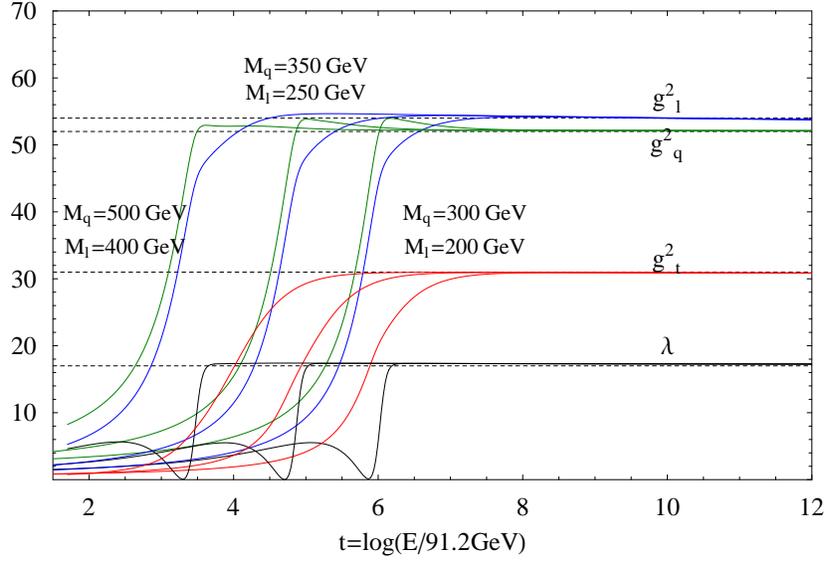}\\
    (b)
    \end{tabular}
\caption{{\small (a) The evolutions of the Higgs quartic and Yukawa couplings for different masses of fourth quarks (from left to right: 250 GeV, 160 GeV, 120 GeV), with fixed mass of the fourth lepton(100 GeV). This is only for illustrative purposes; (b) The evolutions of the Higgs quartic and Yukawa couplings with large masses of the fourth family. The dotted lines represent the roots listed in (\ref{fpvalues}) by solving the $\beta_Y=0$ equations.}}
\label{Yukawa}
\end{figure}

In Fig.\ref{Yukawa} we give the evolutions of the quartic and Yukawa couplings with different masses for the fourth generation quarks. In Fig.\ref{Yukawa}(a) the mass of the fourth generation leptons are taken to be $\overline{m}_l \sim$ 100 GeV, while the fourth quark masses $\overline{m}_q $ are 120 GeV, 160 GeV and 250 GeV respectively. As mentioned at the beginning of this section, these cases are presented only for illustrative purposes in order to show how high the values of $\Lambda_{FP}$ are for the hypothetical light mass case. In Fig.\ref{Yukawa}(b) three heavier mass combinations are plotted. We see that as the initial fourth quark masses increase, the scale $\Lambda_{FP}$ becomes closer to the electroweak scale, from the order of $\sim 10^2 $ TeV to the order of $\sim 10$ TeV. For the $\overline{m}_q \geqslant 400 $ GeV case, the fast-growing region of the couplings even approaches an order of TeV. As in the given example of the heavy case, the fixed point RG values coincide with the roots found by solving $\beta_Y =0 $. This means that the quasi fixed point values of $\lambda$ and the Yukawa couplings are determined merely by the structure of the two-loop RGEs, which is simply characterized by the vanishing $\beta$-functions(\ref{fixedpoint}). Evolutions of the gauge couplings only contribute negligible fluctuations.
The physical consequences of moving the scale $\Lambda_{FP} $ down to TeV level, such as providing an alternative approach to the 
hierarchy problem, will be discussed in a separate paper \cite{HX}. 
\begin{figure}[!tbp]
\centering
    \begin{tabular}{c}
    \includegraphics[scale=1.2]{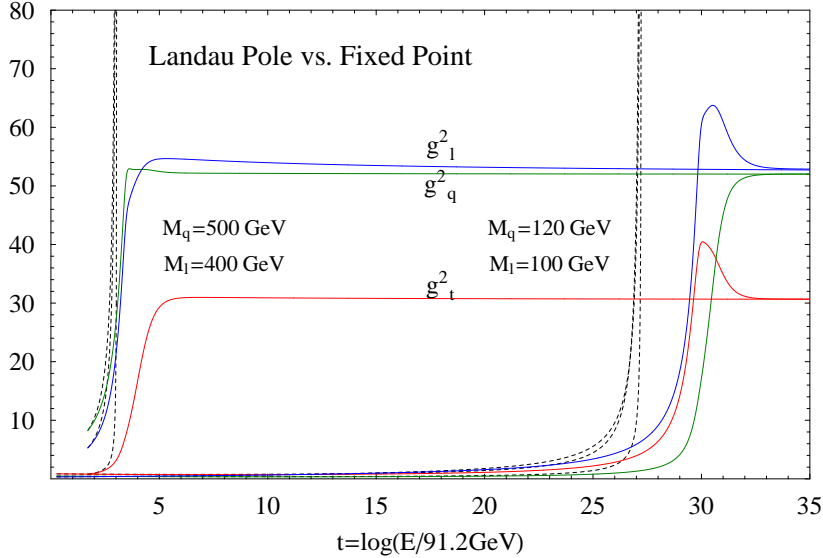}
    \end{tabular}
\caption{{\small The Landau pole(dotted lines) and the quasi fixed point(solid lines) of the Yukawa couplings of the
fourth generation fermions and the top quark. For a heavy fourth generation (left side), both the Landau singularity from
one-loop RGEs  and the quasi fixed point from two-loop RGEs appear at about 2 $\sim$ 3 TeV, while for a light fourth 
generation (right side), their locations at the energy scale differ by two orders of magnitude.  
}}
\label{Landau}
\end{figure}

To further appreciate the significance of  $\Lambda_{FP} $, we show in Fig.\ref{Landau} the evolution of the couplings at one and two loops. It is interesting
to notice that the Landau singularities which appear at one loop are located practically at $\Lambda_{FP} $ for a heavy fourth generation but at
a different energy scale for the light case (which is ruled out experimentally anyway).  The inclusion of two-loop terms transforms the Landau singularity
into a quasi fixed point! 
The significance of $\Lambda_{FP} $ as a physical cut-off scale
in this scenario cannot be underestimated. As we have mentioned above, the dynamics of condensates is controlled by the region of the
Yukawa couplings which is below $\Lambda_{FP} $, i.e. at values of the couplings smaller than those at the quasi fixed point obtained at two loops.
What the quasi fixed point shows was the possibility that scale symmetry is restored  at and above $\Lambda_{FP} $ i.e. in the TeV scales.
\begin{figure}[!tbp]
\centering
    \begin{tabular}{c}
    \includegraphics[scale=1.2]{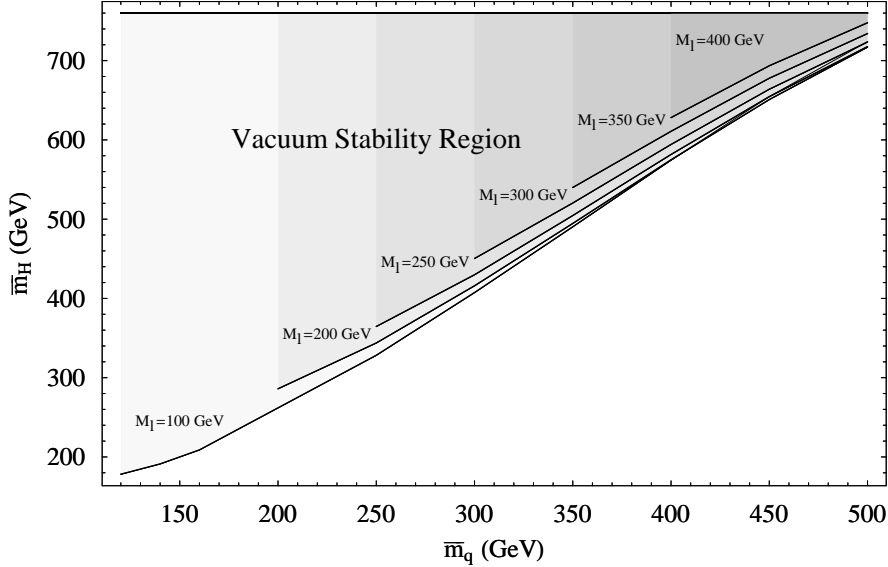}
    \end{tabular}
\caption{{\small The relation between the masses of Higgs and the fourth generation quark. The fourth lepton mass $\overline{m}_l$  is taken to be 100 GeV, 200 GeV, 250 GeV, 300 GeV, 350 GeV and 400 GeV respectively (for the meaning of these masses, note the remarks made after Eq.(\ref{fpmass})). }}
\label{HQ}
\end{figure}

So far we have considered the fourth quark/lepton masses and gauge couplings, what about the Higgs mass at the electroweak scale? 
In fact in Figs.\ref{light_heavy} and \ref{Yukawa} we have considered a lower bound for $\overline{m}_H$, by imposing the vacuum stability
condition \cite{stability} such that the running of the quartic coupling $\lambda$ is always greater than or equal to zero. For the examples
shown in Figs.\ref{light_heavy} and \ref{Yukawa}, the initial value of $\lambda$ is the minimal one to ensure the vacuum stability. In other
words, $\lambda$ can reach zero ($\lambda_{dip} \approx 0$) if we adjust its initial value. This can be seen from the ``dip" of the RG
trajectory of the quartic coupling $\lambda$ shown in Figs.\ref{light_heavy} and \ref{Yukawa}. 
An increase in the initial values of $\lambda$ also increases $\lambda_{dip}$, 
but it will not change the pattern of the running of the Yukawa and quartic couplings and the location of the fixed point
$\Lambda_{FP}$. The physical meaning of this ``dip" is related to the condensation of fourth generation fermions and will be addressed
in details in the next section. For the light mass case, $\lambda_{dip}$ is very sensitive to the initial value of $\lambda$ and
there is a fine-tuning problem if we adjust the initial values of $\lambda$
for $\lambda_{dip} \approx 0$. Taking the case $m_q=120 $ GeV, $m_l=100$ GeV as an example, for $\lambda_{dip}$ to be positive and close to zero, say  $ 0<\lambda_{dip}<0.5$, one has to fine-tune the initial value of $\lambda$ to at least eight decimal places ($\lambda=0.26194035$). For the cases $m_q=250$ GeV, $m_l=100$ GeV and $m_q=400$ GeV, $m_l=100$ GeV, 
 one only needs to tune four ($\lambda=0.8890$) and two decimal places ($\lambda=2.72$) respectively, for
 $\lambda_{dip}$ to fall into the same range. Therefore the heavy mass case is also favoured by the condition $\lambda_{dip} \approx 0$, since much less fine-tuning is needed for the initial value of $\lambda$.  
In Fig.\ref{HQ} we take the fourth lepton mass $m_l$ to be 100 GeV, 200 GeV, 250 GeV, 300 GeV, 350 GeV and 400 GeV respectively and plot the mass relation between Higgs and the fourth quarks. Note that these masses are calculated naively as mentioned in the remarks made after 
Eq.(\ref{fpmass}). The lower bounds of the Higgs mass in Fig.\ref{HQ} are determined from the minimal values of the initial $\lambda$ and they fall into a narrow strip, which may be called the ``condensate band" whose meaning will be clarified later.  The different fourth lepton masses do not affect this bound much. A change of mass
$\sim$ 250 GeV of the fourth lepton only causes a change of $20 \sim 30 $ GeV in the lower bound of the Higgs mass.
With this uncertainty we can describe the area under the strip as excluded values of the Higgs mass. For example, a 350 GeV fourth quark needs the Higgs mass to be above $500 \pm 20$ GeV as it can be seen from Fig.\ref{HQ}.

\section{Bound States and Condensates of the Fourth Generation}

~~~~

In the previous section, the evolutions of the RGEs of the Higgs quartic and Yukawa couplings, with large initial masses at electroweak scale, have shown that, close to the scale $\Lambda_{FP} \sim$ TeV, these couplings increase rapidly ($\lambda$ has a couple of turning points) and then run into a fixed point in the strong Yukawa coupling region. If we consider the scales between the electroweak scale $\Lambda_{EW}$ and the new scale $\Lambda_{FP}$ as the perturbation region, usually we need specify the boundary conditions at either scale or both, and then run the RGEs bottom-up (IR to UV) or top-down (UV to IR) depending on which boundary conditions are imposed.
First we recall the {\it compositeness conditions} which is related to the top/fourth quark condensation or topcolor-assisted
technicolor models \cite{Hill1990, Frampton, Holdom:2006}.In these models, the Higgs quartic and Yukawa couplings are assumed to become divergent when approaching the Landau pole \cite{Hill1990, hung2},
\be \label{comp1}
\lambda(t), ~g_f (t) \longrightarrow \infty,  ~\textrm{when}~~t \rightarrow \Lambda_{L},
\ee
\be \label{comp2}
\lambda(t)/g^2_f (t) \longrightarrow \textrm{const.},  ~\textrm{when}~~t \rightarrow \Lambda_{L}.
\ee
The result from a one-loop calculation \cite{hung2} satisfies these conditions
(\ref{comp1}) and (\ref{comp2}), while the two-loop result in this paper shows that the quartic and Yukawa couplings are not divergent but run into some quasi fixed point respectively,i.e., the Landau pole is replaced by the fixed point and the compositeness
condition is replaced by the {\it fixed-point} behavior
\be \label{comp3}
\lambda(t), ~g_f (t) \longrightarrow \textrm{const.},  ~\textrm{when}~~t \rightarrow \Lambda_{FP}
\ee
In both cases the quartic and Yukawa interactions become strong above $\Lambda_{FP}$ and the inclusion of higher loop contributions is beyond
the scope of this paper. As mentioned in the previous section, we assume that the behavior shown in Eq. (\ref{comp3}) still holds to higher
loop level, and the higher loop contributions will not shift the location of $\Lambda_{FP}$ and modify the values of the quartic and Yukawa
couplings at the fixed point significantly. 

An important point that was pointed out in the previous sections is the following. The dynamics of electroweak-symmetry breaking condensates
is effective at values of the Yukawa couplings which are smaller \cite{HX2} than those at the quasi fixed point. In particular,
the "dip"  in the quartic coupling shown in Fig. (\ref{condensates}) corresponds to the case where the short-range Yukawa potential becomes a 
long-range Coulomb-like potential with strong Yukawa couplings, albeit those with values smaller than the fixed point values. We shall come
back to this point below. For the moment, let us take at face value the quasi fixed point values of the couplings and investigate the
kind of bound states that could or could not be formed with this assumption. 

Here we restrict ourselves to the two-loop level and focus on the quasi fixed point behavior (\ref{comp3}).
As the Yukawa couplings of the fourth generation become strong around the scale $\Lambda_{FP}$ and above, it is possible for the heavy quarks or leptons to form bound states. The fixed point from the point of view of RG flow, might correspond to a critical coupling of the bound states, similar to the critical couplings studied in QED/QCD with dynamical symmetry breaking, e.g. $\alpha^{\textrm{QED}}_c = \pi/3$ \cite{Miranski}.
In fact a full relativistic treatment in \cite{DESB_HX} shows that there does exist a critial point in the Yukawa coupling of the Standard Model and the critical coupling is $\alpha_c^{\textrm{Yukawa}} = \pi/2 \approx 1.57$. We have made some comments in Section II concerning the value of the Yukawa coupling where condensate formation can occur i.e. $\alpha_{q,l} > \alpha_c$, and this could be a factor of 3 smaller than the quasi fixed point values (\ref{fpvalues2}). Also notice that when  $\alpha_{q,l} > \alpha_c$ is satisfied, the corresponding value for the top quark is still too small to form a condensate. For heuristic reasons, we will use a simple-minded non-relativistic approach in this manuscript to investigate bound state formation. As can be seen below, this approach also confirms our relativistic expectation: The top quark cannot form a bound state.

In the present section we perform an analysis at the non-relativistic limit, following \cite{HungRho}. This means
that we will solve the Schr\"{o}dinger equation to find the condition for bound state formation
and to estimate the binding energy. A careful examination of Fig. (2)
reveals two interesting regions around $\Lambda_{FP}$. Region I: This is where the quartic coupling has a ``dip''
just before the energy scale where it reaches its fixed point value. Region II is the fixed-point region.
This is shown in Fig.(\ref{condensates}) for the case $m_q= 450$ GeV and $m_l=350$ GeV (other heavy mass cases have the similar figures).
\begin{figure}[!tbp]
\centering
    \begin{tabular}{cc}
    \includegraphics[scale=0.77]{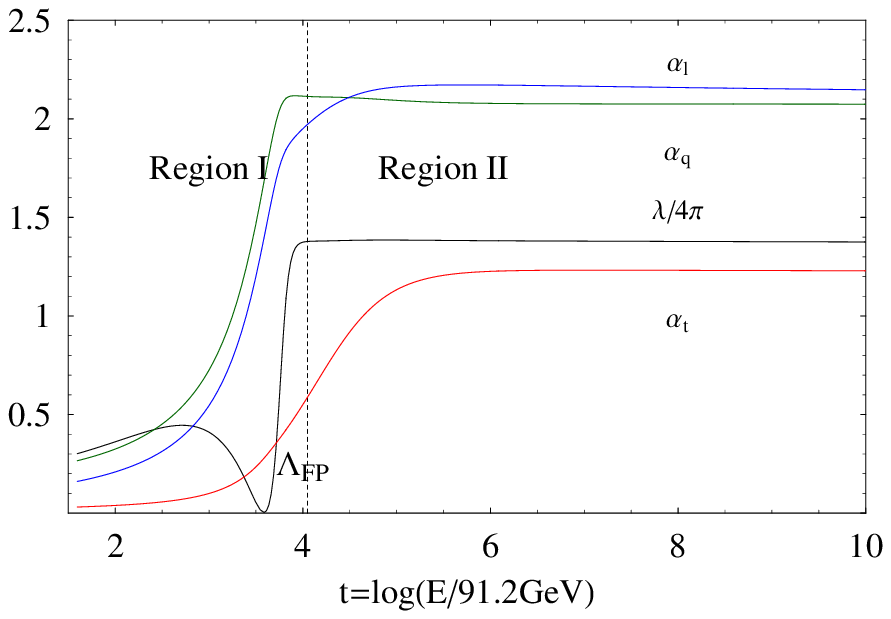} & \includegraphics[scale=0.77]{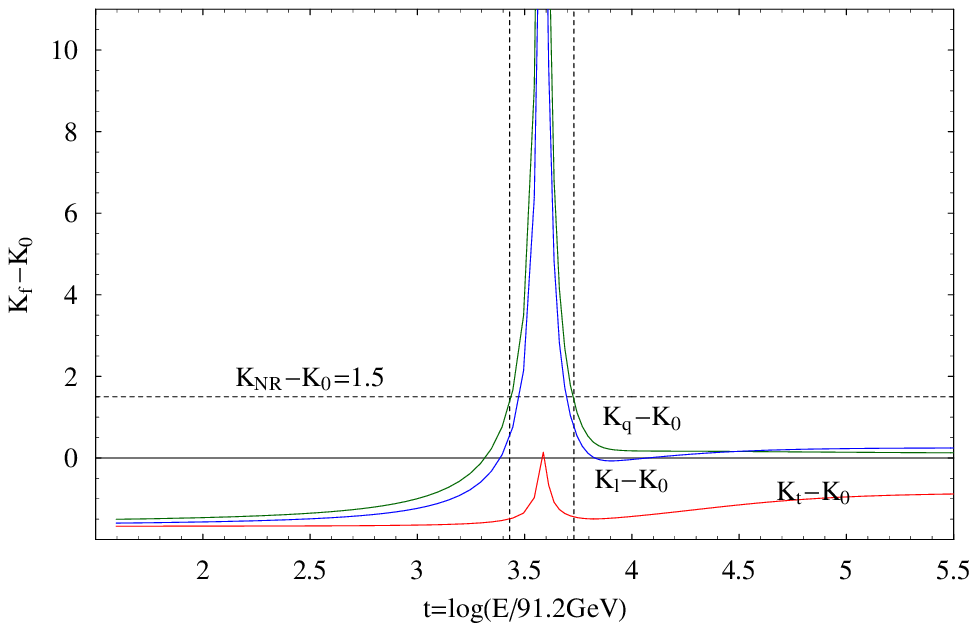}\\
    (a) & (b)
    \end{tabular}
\caption{{\small ($m_q=450$ GeV and $m_l=350$ GeV)~(a) The locations of $\Lambda_{FP}$, Region I and II; (b) $K_f - K_0$ with $ K_f = g_f^3 / 16 \pi \sqrt{\lambda}$ and $K_0=1.68$. For illustration purpose, the initial value of $\lambda$ is increased slightly such that the peak value of $K_f$ would not become too large to fit in the figure. The horizontal dotted line indicates an estimate of $K_f$ where the non-relativistic method is still applicable and the vertical dotted lines enclose the region where a fully relativistic approach is needed.}}
\label{condensates}
\end{figure}
Region I where the dip occurs would give rise to either a Coulomb-like potential with strong coupling between the heavy
fermions or a Yukawa potential with a ``small Higgs mass''. In this region, tight bound states of fourth generation fermions
can be formed. It is, however, beyond the scope of this paper to fully discuss this
interesting region and it will be presented in \cite{HX2}. However, we will sketch what we might expect from that region.
We will discuss in more details bound state formation in Region II, first using
a heuristic variational method, followed by a more accurate numerical solution to the Schr\"{o}dinger equation.

The non-relativistic Higgs-exchange potential is given by
\be
\label{yukpot}
V(r) = - \alpha_Y(r) \frac{e^{-m_H(r) r}} {r}
\ee
where $m_H$ is the Higgs mass and $\alpha_Y = \frac{m_1 m_2 } {4\pi v^2} $ with $v=246$ GeV. The masses of two fermions are $m_1 $ and $m_2$ respectively and the reduced masses of the system is $ M = m_1 m_2 /(m_1 +m_2)$. The total energy of the system is $ E_{tot}$ and the energy of the center of mass is $ E_{cm}$.
For simplicity we consider the state $l=0$. To gain an insight into conditions for bound state formation, we start out
with the Rayleigh-Ritz variational method for pedagogical purpose. This will be followed by a more accurate numerical treatment.

In Eq. (\ref{yukpot}), the $r$-dependence (or equivalently the energy dependence) of $\alpha_Y$ and $m_H$ is written down for
the following reasons. In Region (II), the quartic and fourth generation Yukawa couplings are fixed for $E> \Lambda_{FP}$ (or
equivalently at ``short distances''). In that region  $\alpha_Y$ and $m_H$ are constants. In Region (I), just below $\Lambda_{FP}$,
the quartic coupling varies rapidly: $m_H$ decreases while the fourth generation masses are approximately constant. This means
that the Yukawa interactions are now of much longer range than in Region (II). This will have important consequences
concerning condensate formation as we will discuss below.

In Region (II), we will take the following fixed point values for $\alpha_Y$ and $m_H$:
$\alpha_Y \approx 2.09, 2.16, 1.22$ for Q, L and t respectively, and $m_H \approx 1.446 $ TeV.
The trial wave function is taken to be $u (y, r)=2 y^{\frac{3}{2}}~e^{-y r}$ and $y$ is the variational parameter \cite{HungRho}.
The relative energy $ E = E_{tot} - E_{cm} $ is given by
\be
E= \frac{\hbar^2}{2 M} y^2 - \frac{4 \alpha_Y y^3}{(2y+m_H)^2}.
\ee
Redefining variables $ z=2y/m_H, K_f= 2M \alpha_Y / (m_H \hbar^2) $ and applying $ dE/dz=0$ yield $ K_f=(1+z)^3/z(z+3) $.
Then the optimum energy is ($\hbar =1$)
\be
E=-\alpha_Y m_H \frac{z^3(z-1)}{4 (z+1)^3}.
\ee
The bound state condition is found by requiring $z>1$ or equivalently
\be
\label{K}
K_f >2, ~~~~(\textrm{variational method})
\ee
where, for simplicity, we specialize to the case $m_1=m_2=m_f$ and $K_f \equiv \alpha_Y m_f/m_H $. In terms of the quartic and Yukawa couplings, $K_f$
can be rewritten as
\be 
K_f = \frac{g_f^3}{16 \pi \sqrt{\lambda}}.
\ee
Note that $K_f$ is independent of the electroweak scale $v$. 
What are the $K_f$ values for Q, L and t?
Using the fixed-point values, we obtain $K_q=1.82$, $K_l=1.92$, and $K_t=0.82$. Comparing $K_q$,
$K_l$ and $K_t$ with the bound state condition (\ref{K}) obtained from the variational method, one
notices that the fourth generation quarks and leptons can marginally satisfy (\ref{K}) while the top
quark certainly cannot. The next question is whether the fourth generation quarks and leptons can actually
form bound states in Region (II).

A numerical solution to the Schr\"{o}dinger equation \cite{Poliatzky} actually yields
\be
\label{Knum}
K_f>1.68, ~~~~(\textrm{numerical method}).
\ee
$K_q$ and $K_l$ indeed satisfy the bound state constraint (\ref{Knum}) which definitely rules out
the formation of a $t\bar{t}$ bound state. However, $K_q=1.82$, $K_l=1.92$ are still close to the lower bound 1.68 in
(\ref{Knum}) which suggests that $Q\bar{Q}$ and $L\bar{L}$ are loosely bound. The binding energy
can be calculated numerically following \cite{Poliatzky}
\be \label{binding}
\sum_{n=0}^{n_{\textrm{max}}} (-K_f)^n \varphi_n (n_{\textrm{max}} +1 -n, \nu) =0
\ee
where $\nu =2\sqrt{-m_f E}/m_H$ and the integer $n_{\textrm{max}}$ is introduced to truncate the series.
The functions $\varphi_n (w, \nu)$ come from the expansion of the wave function and satisfy the recurrence relation
\bea \label{recurrence}
\nonumber
\varphi_0 (w, \nu)&=&1 \\
\varphi_n (w, \nu)&=& \int_0^w dx \frac{[(x+n-1)(x+n-1+\nu)]^l}{[(x+n)(x+n+\nu)]^{l+1}}\varphi_{n-1} (x, \nu)  ~~~~~n=1, 2, ...
\eea
In our case the angular momentum $l=0$. For $K_q=1.82$, one can now solve (\ref{binding}) which yields $\nu =0.108$.
This gives a rather small binding energy $E_q \approx -4.9 $ GeV. For the lepton with $K_l=1.92$, we obtain $\nu=0.196$ and
$E_l \approx - 15.7 $ GeV. This confirms the expectation that $Q\bar{Q}$ and $L\bar{L}$ in region (II) are loosely bound.
Note that the Eq. (\ref{binding}) has  an expansion parameter $K_f \approx 2 $, but is still a fast-converging series. For example,
the critical value of $K_f $ is determined by solving Eq. (\ref{binding}) for zero binding energy $ \nu=0$. In \cite{Poliatzky}
this value was found to be $ K_f = 1.6798077 $ for $ n_{\textrm{max}} = 20$, however, if one just takes $n_{\textrm{max}}=3$,
then $K_f = 1.6803$ and even for  $n_{\textrm{max}}=2$, $K_f = 1.696$, which is quite close to the result which comes from
the summation of the power series of $K_f$ up to $ K_f^{~20} $. Similar to Wilson and Fisher's example of the 
critical exponent calculation in Sec. II, we see that the lower orders of expansion of $K_f$ are already good enough to give
an accurate value.

What happens when one moves from Region (II) into Region (I)? Just below $\Lambda_{FP}$, the quartic coupling decreases
rapidly as shown in Fig. (\ref{condensates}). This means that the Yukawa interactions become increasingly
long-range. This reminds us of the Landau theory of phase transition where the correlation function
between two Ising spins takes the form $G(r) = \exp(-r/\xi)/4\pi r$, where $\xi$ is the correlation length. (In our case
this would be $\xi_{H} \sim 1/m_H $.)
In a nutshell, the phase transition occurs when one goes from a short-range correlation (small $\xi$) to
an infinite-range correlation ($\xi= \infty$).

Moving into Region (I) with a changing $m_H$ and approximating the fourth generation Yukawa couplings to
be constant as shown in Fig. (\ref{condensates}), one can plot $K_f -K_0$ ($K_0=1.68$) versus $t$ for
the fourth generation quarks and leptons as well as for the top quark. $K_f -K_0=0$ corresponds to
bound states that are barely formed, with anything above that corresponding to tighter bound states.
One also notices that $K_f -K_0$ for the top quark is always below the $K_f -K_0=0$ line which
implies that there can be at most loosely bound states $t \bar{t}$, but no condensates.
When the binding energy is comparable in magnitude to the mass of the constituent fermions, our non-relativistic
approximation breaks down. 
(see \cite{DESB_HX} for a relativistic analysis.)
This corresponds to the dashed line in the second figure of
Fig. (\ref{condensates}). Notice that Fig. (\ref{condensates}) 
is for the case when $m_H$ at the dip is small but finite. This depends on the initial value of $m_H$
at the electroweak scale. However, vacuum stability only requires that $\lambda \geq 0$ and hence the ``dip''
could correspond to the point where $\lambda$ {\em vanishes}. This happens in a region which is very close to
$\Lambda_{FP}$ and, as a result, the Yukawa couplings of the fourth generation can be considered to be nearly constant.
As we have mentioned above, when one goes from Region (II) above $\Lambda_{FP}$ to Region (I),
the interaction induced by the exchange of the Higgs scalar becomes increasingly long-range until this range
becomes infinite at the ``dip''. One may expect some kind of phase transition due to the
formation of Higgs-like condensates. Heuristically, at the ``dip'',
the Yukawa potential becomes effectively an attractive Coulomb-like potential of the form
\be
\label{coulomb}
V_{\textrm{``dip''}}(r) = - \frac{\alpha_Y} {r} \,,
\ee
where, e.g., $\alpha_Y \approx 1.6$ (at the "dip") for a $Q\bar{Q}$ system. This represents a strong Coulomb-like potential.
Studies of condensed matter systems \cite{CDM} suggest such a potential
could potentially lead to a formation of condensates. In our case, with (\ref{coulomb}), one might expect
condensates of the type $\langle \bar{Q}_L Q_R \rangle$ to get formed and to play the role of a dynamical
Higgs condensate.  This is what we refer to above as additional (dynamical) Higgs doublets which can
can contribute to the symmetry breaking of the Standard Model. It is however beyond the scope of this
paper to discuss this interesting issue at length \cite{DESB_HX, HX}. It will be treated in \cite{HX2}.

Another important aspect of the formation of Higgs-like condensates with the fourth generation quarks
and leptons is the value of the condensates. One expects that $\langle \bar{Q}_L Q_R \rangle$ and
$\langle \bar{L}_L L_R \rangle$ to be proportional to the scale where the Yukawa couplings grow strong
for condensates to form (similar to dynamical symmetry breaking in Technicolor models e.g.). From
Fig. (2) and Fig. (\ref{condensates}), one can see that this is around $\Lambda_{FP}$. As a result,
one might expect e.g. $\langle \bar{Q}_L Q_R \rangle \sim - c \Lambda_{FP}^3$ where $c$ is a constant
which depends on the details of the dynamics. These condensates would contribute to the breaking
of the SM and, as a result, are bounded in values: They should not exceed the electroweak scale.
As we can see from Fig. (2), $\Lambda_{FP}$ varies from O(TeV) to some GUT scale value depending on whether
the fourth generation is ``heavy'' or ``light''. This implies that one has to ``fine tune'' the dynamics
more and more accurately as $\Lambda_{FP}$ increases in order to keep the condensate values close
to the electroweak scale. This ``fine tuning'' issue and its connection with the hierarchy problem
are discussed in \cite{DESB_HX, HX}.

It is interesting to write down an effective action which can incorporate the aforementioned bound states (with arbitrary spin).
In particular, we would be interested in the effects of dynamical Higgs-like scalar bound states on the effective
potential \cite{HX2}. This hybrid (fundamental plus dynamical) Higgs spectrum will provide a rich phenomenology involving
couplings to fermions which are, in principle, calculable and testable \cite{HX2}.

\section{Conclusions and Discussions}

~~~

A heavy fourth generation of quarks and leptons is found to lead to an interesting physical scenario as discussed in this paper. From the two-loop RGE running of the Higgs quartic and Yukawa couplings and gauge couplings, we have shown the existence of a quasi fixed point of the quartic and Yukawa couplings of the top quark and the fourth family quarks and leptons.
The quasi fixed point masses are found to be
$
\overline{m}_H^{*} = 1.45 ~\textrm{TeV}, \overline{m}_t^{*} = 0.97 ~\textrm{TeV}, 
\overline{m}_q^{*} = 1.26 ~\textrm{TeV}, \overline{m}_l^{*} = 1.28 ~\textrm{TeV}.
$
The quasi  fixed point scale $\Lambda_{FP}$ ranges from a few TeV to the order of $10^2$ TeV, depending on the masses of the fourth generation at electroweak scale $\Lambda_{EW}$. As we have mentioned above, the dynamics of condensates is controlled by the region of the
Yukawa couplings which is below $\Lambda_{FP} $, i.e. at values of the couplings smaller than those at the quasi fixed point obtained at two loops.
What the quasi fixed point shows was the possibility that scale symmetry is restored  at and above $\Lambda_{FP} $ i.e. in the TeV scales. Based on the observation of the existence of a two-loop quasi fixed point at an energy scale of O(TeV), we {\em conjecture} that a true fixed point exists a similar scale, albeit with possibly a lower value for the Yukawa couplings. A proof of this conjecture is beyond the scope of this paper.

When Yukawa interactions become strong, bound states of the fourth generation can be formed at TeV scales, while
the top quark can hardly do so as we have seen in Sec. III. The formation of these fourth generation bound states
has far-reaching consequences concerning the Higgs-like condensates such as $\langle \bar{Q}_L Q_R \rangle$, 
$\langle \bar{L}_L L_R \rangle$ and the bound states with various spins of the form $\bar{Q} Q$, $\bar{L} L$ and even 
``leptoquarks'' $\bar{Q}L$ +H.c. due to the strong Yukawa interactions near the ``dip''. 
As described in Sec. III and looking, in particular, at Fig. (\ref{condensates}),
the fourth generation quarks and leptons are loosely bound above the fixed point scale $\Lambda_{FP}$. The
correlation length $\xi_{H} \sim 1/m_H $ which is small above $\Lambda_{FP}$ becomes increasingly large as the
energy decreases below $\Lambda_{FP}$ until it becomes infinite at the ``dip'' shown in Fig. (\ref{condensates}).
This behaviour is indicative of a phase transition with the formation of condensates whose implications
are discussed in \cite{DESB_HX, HX}. This behaviour appears to be independent of the location
of $\Lambda_{FP}$ as one can see from Fig. (2). However, the higher the value of $\Lambda_{FP}$ is, the more ``fine-tuning''
is needed in order to maintain this ``dip'' at $\lambda \geq 0$. As we have mentioned in Sec. III, one expects
the values of the condensates to be at the scale where the Yukawa couplings are strong enough for their formation
i.e. at around $\Lambda_{FP}$. One expects e.g. $\langle \bar{Q}_L Q_R \rangle \sim - c \Lambda_{FP}^3$ where $c$ is a constant
which depends on the details of the dynamics. Since these condensates spontaneously break the electroweak symmetry, they are
required to have values not exceeding the electroweak scale. As we can see from Fig. (2), $\Lambda_{FP}$ varies from O(TeV)
to some GUT scale value depending on whether
the fourth generation is ``heavy'' or ``light''. This implies that one has to ``fine tune'' the dynamics
more and more accurately as $\Lambda_{FP}$ increases in order to keep the condensate values close
to the electroweak scale. For the gauge couplings, their evolutions may change significantly when new dynamical 
degrees of freedom emerge at $\Lambda_{FP}$. This and other issues might have interesting consequences
concerning the hierarchy problem as discussed in \cite{DESB_HX, HX} although much is needed to actually complete
the proof.

The aforementioned bound states may work as additional Higgs doublets in two or three Higgs models.
The hybrid (fundamental plus dynamical) Higgs mass spectrum will give rise to a rich phenomenology which will be relevant to the LHC 
and even to the ILC \cite{HX2}. For the fourth quarks themselves, the large mass $m_q \gtrsim 400$ GeV seems to be favoured if
one wishes to have a TeV-scale $\Lambda_{FP}$ and if one considers the ``vacuum stability naturalness" issue (Sec. II). 

\section*{Acknowledgments}

We would like to thank George W. S. Hou and Hank Thacker for the discussions.
This work is supported in parts by the US Department of Energy under grant No. DE-FG02-97ER41027.

\end{document}